\documentclass[showpacs,preprintnumbers,amsmath,amssymb]{revtex4-1}
\usepackage{amssymb}
\usepackage{amsmath}
\usepackage{epsfig}
\usepackage{graphicx}
\renewcommand{\vec}[1]{\mbox{\boldmath$#1$}}
\newcommand{\be}{\begin{equation}}
\newcommand{\ee}{\end{equation}}
\newcommand{\bea}{\begin{eqnarray}}
\newcommand{\eea}{\end{eqnarray}}
\numberwithin{equation}{section}
\begin{document}
%\begin{frontmatter}
\title{Path integrals for dimerized quantum spin systems}
\author{Adriana Foussats}
%A.\ Greco, and A.\ Muramatsu}, 
\author{Andr\'es Greco}
\affiliation{Facultad de Ciencias Exactas, Ingenier\'{\i}a y Agrimensura and 
Instituto de F\'{\i}sica Rosario
(UNR-CONICET). 
Av. Pellegrini 250-2000 Rosario, Argentina}
\author{Alejandro Muramatsu}
\affiliation{Institut f\"ur Theoretische Physik III, Universit\"at Stuttgart,
Pfaffenwaldring 57, D-70550 Stuttgart, Germany}

\begin{abstract}
Dimerized quantum spin systems may appear under several circumstances, e.g\ by a modulation
of the antiferromagnetic exchange coupling in space, or in frustrated quantum antiferromagnets.
 In general, such systems display a quantum phase transition to a N\'eel state as a function of a
 suitable coupling constant. We present here two path-integral formulations appropriate for spin $S=1/2$
 dimerized systems. The first one deals with a description of the dimers degrees of freedom in an SO(4)
 manifold, while the second one provides a path-integral for the bond-operators introduced 
by Sachdev and Bhatt. The path-integral quantization is performed using the Faddeev-Jackiw symplectic 
formalism for constrained systems, such that the measures and constraints that result from the algebra
 of the operators is provided in both cases. As an example we consider a spin-Peierls chain, 
and show how to arrive at the corresponding field-theory, starting with both an SO(4) formulation 
and bond-operators.    
\end{abstract}

\maketitle

\vspace*{0.5cm}

%\end{frontmatter}
\section{Introduction}
Dimerized quantum spin systems appear frequently in quantum antiferromagnets, 
leading to a spin-liquid in the form of a valence-bond solid (VBS) for strong enough dimerization. 
A well known example is given by spin-Peierls systems, where in quasi one-dimensional materials the dimerization can arise due to electron-phonon interaction \cite{cross79,uhrig97}. Dimerized systems constitute also a starting point to study frustrated quantum antiferromagnets in two dimensions. In such a way, models with antiferromagnetic exchange interactions beyond nearest neighbors were addressed \cite{gelfand89}, finding spin-Peierls states, as were suggested from topological events in SU(N) extensions of quantum antiferromagnets \cite{read89,read90}. Dimerized states also result from frustrating longer-range interactions as in the case of the Shastry-Sutherland model \cite{shastry81}, where nearest neighbor ($J$) and next-nearest neighbor ($J'$) interactions on alternating plaquettes lead to a VBS for large enough values of $J'/J$ ($J'/J \geq 1$ in the isotropic case for spin $S=1/2$ and in the absence of a magnetic field). In all the cases above, a quantum phase transition (QPT) between a N\'eel state or some other intermediate state, and a VBS sets in by varying the ratio of the competing couplings. In particular, in the case of the Shastry-Sutherland model, several intermediate phases between the N\'eel state and the VBS were proposed
\cite{muellerhartmann00,koga00,chung01,laeuchli02}. However, the nature of the intermediate phase is still not clear \cite{knetter00,takushima01,carpentier01,zheng01}.

Two-dimensional dimerized quantum antiferromagnets were also discussed recently \cite{yoshioka04,wenzel08}
in connection to the proposal of deconfined quantum criticality \cite{senthil04}. 
In this theory, the critical point is characterized by the deconfinement of new degrees 
of freedom resulting from fractionalization of the order parameter, as opposed to the
 Landau-Ginzburg-Wilson (LGW) paradigm of phase transitions, where the critical point 
is characterized solely by the critical behavior of the order parameter and its correlators.
 Specifically, the theory by Senthil {\em et al.} \cite{senthil04} allows for a continuous phase
transition between a N\'eel state, that breaks rotation symmetry in spin space,
and a VBS, that spontaneously breaks the lattice symmetry, while on the grounds of an LGW-theory,
 a first order phase transition is generally expected. The quantum Monte Carlo simulations in Ref.\ \cite{wenzel08}
 show for different dimerization patterns of an $S=1/2$ antiferromagnetic Heisenberg model on the square lattice that,
 while most of the dimerization patterns lead to critical exponents consistent with the universality class of the 
three-dimensional O(3) Heisenberg model, a staggered pattern leads to deviations from it. Hence, the authors raised 
the possibility of deconfined quantum criticality in this case. 

A QPT from a N\'eel to a paramagnetic phase can be approached from the magnetically ordered side by introducing
 spin coherent states \cite{perelomov86}, that lead in general to an O(3) non-linear $\sigma$-model \cite{zinnjustin90,sachdev99}, 
as in the treatment of the antiferromagnetic quantum Heisenberg model in the context of high temperature
 superconductivity \cite{chakravarty89}, or, alternatively, employing a CP$^1$ representation \cite{yoshioka04}. 
However, the discussion above, shows that it would be certainly interesting to address the QPT from the VBS side
 in the case of a dimerized system. When the VBS consists of nearest neighbor dimers, for spins $S = 1/2$, 
the dimers are naturally described by the possible states of two spins, i.e.\ by the manifold SU(2) $\otimes$ SU(2) $\simeq$ SO(4). 
Some time ago, bond-operators were introduced \cite{chubukov89,sachdev90,chubukov91} for that purpose, that offer an intuitive picture
 for singlets and triplets. The application of such a representation with different approximation schemes, mostly based on mean-field
 theory led to rather good descriptions of spin-liquid states in a variety of situations in frustrated \cite{sachdev90,chubukov91}, 
dimerized \cite{kotov98,matsumoto02}, and bilayer antiferromagnets \cite{vojta99}. The aim of the present work is to formulate 
a path integral dealing directly with the SO(4) manifold on the one hand, and with bond-operators on the other hand, enabling a
 field-theoretic treatment of the transition between dimerized and other possible states.

We consider here two equivalent approaches. The first one deals with the generators of the 
SO(4) algebra. While previous work \cite{vanduin97} used coherent states, we perform the path-integral
quantization using the Faddeev-Jackiw symplectic formalism for quantum field-theories with constraints 
\cite{Faddeev88,Govaerts90}. The treatment of constrained systems was initiated by Dirac \cite{Dirac64}
 and continued by Faddeev and Jackiw \cite{Faddeev88}. These  methods were used before for the Heisenberg and
 $t-J$ model \cite{foussats99a,foussats99b,foussats00,falb08} showing consistent results with other works
 were coherent states were used \cite{wiegmann88}. The formal development is shown in Sec.\ \ref{SO4}. 
The second approach, based on a path-integral representation of bond-operators, is described in Sec.\ \ref{bondF}. 
In both cases, special emphasis is given to the determination of the measures and constraints appropriate for
 the respective algebras. We expect in this way to allow for the treatment of fluctuations beyond the mean-field 
approximation, at the same level as in previous treatments leading to the O(3) non-linear $\sigma$-model starting
 from a N\'eel configuration. Section \ref{SPchain} illustrates how the obtained path integral representations may 
be used to reach a field-theory for a spin-Peierls chain. This example was chosen because the effective field-theory 
is characterized by the presence of a topological term that ensures that in the absence of dimerization, the spin-gap closes. 
 In Appendix A we show explicitly, how to transform from one representation into the other.

\section{Path integral formulation for SO(4) fields \label{SO4}}
\subsection{SO(4) operators for a bond\label{SO4Algebra}}
We start by considering the four states $| \mu \rangle$ for a bond joining two $S=1/2$
states, 
\bea
\mid 0 \rangle & = & \frac{1}{\sqrt{2}}
\left( \mid \uparrow \downarrow \rangle - \mid  \downarrow \uparrow \rangle \right) \;, 
\nonumber \\  
\mid 1 \rangle & = & -\frac{1}{\sqrt{2}}
\left( \mid \uparrow \uparrow \rangle - \mid \downarrow \downarrow \rangle \right) 
\; ,
\nonumber \\
\mid 2 \rangle & = & \frac{i}{\sqrt{2}}
\left( \mid \uparrow \uparrow \rangle + \mid \downarrow \downarrow \rangle \right) 
\; ,
\nonumber \\
\mid 3 \rangle & = & \frac{1}{\sqrt{2}} \left( \mid \uparrow
\downarrow \rangle + \mid  \downarrow \uparrow \rangle \right) \; .
\eea 
We can introduce the Hubbard or  $X$-operators \cite{Hubbard63}
defined as 
\bea \label{OpHubb} 
{\hat X}^{\mu \nu} \equiv \mid \mu \rangle \langle \nu \mid \; , 
\eea 
with $\mu=0,\dots,3$. Trivially, they obey the following commutation rules: 
\bea \label{CommutatorX}
\left[{\hat X}^{\mu \nu} , {\hat X}^{\lambda \rho} \right] & = & {\hat X}^{\mu \rho}
\delta_{\nu \lambda} - {\hat X}^{\lambda \nu} \delta_{\rho \mu} \;  
\eea
and the completeness relation 
\bea \label{CompletnessX}
\sum_\mu {\hat X}^{\mu \mu} = 1 \; . 
\eea 
Since we are considering states of SU(2) $\otimes$ SU(2) $\simeq$ SO(4), we have to construct the six generators of SO(4) out of the 16 operators ${\hat X}^{\mu \nu}$. These generators can be written as 
\bea
\label{ST_operators} 
{\hat {\cal T}}^a & \equiv & -i \varepsilon^{abc} {\hat X}^{bc} \; ,
\nonumber\\
{\hat {\cal S}}^a & \equiv & {\hat X}^{0 a} + {\hat X}^{a 0} \; , 
\eea 
where the Latin indices run over 1,2,3. Using the commutation relations (\ref{CommutatorX}) we obtain 
\bea 
\label{CommutatorTS}
\left[{\hat {\cal T}}^a , {\hat {\cal T}}^b \right] & = & i \varepsilon^{abc} {\hat {\cal T}}^c \; ,
\nonumber\\
\left[{\hat {\cal T}}^a , {\hat {\cal S}}^b \right] & = & i \varepsilon^{abc} {\hat {\cal S}}^c \; ,
\nonumber\\
\left[{\hat {\cal S}}^a , {\hat {\cal S}}^b \right] & = & i \varepsilon^{abc} {\hat {\cal T}}^c \; ,
\eea 
the commutation relations of the generators of SO(4) \cite{biedenharn61}.

The generators of SU(2) for each site of the bond can be also constructed in the following way:
\bea
\label{SpinFromTS}
{\hat S}^a_{(1)} & = & \frac{1}{2} \left({\hat {\cal T}}^a + {\hat {\cal S}}^a \right) \; ,
\nonumber \\
{\hat S}^a_{(2)} & = & \frac{1}{2} \left({\hat {\cal T}}^a - {\hat {\cal S}}^a \right) \; ,
\eea
where 1 and 2 denote the two spins making up the bond, such that any spin Hamiltonian can be expressed in terms of the generators of SO(4).

Finally, we consider the Casimir operator in SO(4). The Casimir operator, which commutes with the generators,  is given by the sum of the squares of the generators \cite{cornwell2} {\it i.e.} 
${\hat {\vec {\cal T}}}^2 + {\hat {\vec {\cal S}}}^2 $. From the relations (\ref{SpinFromTS}), we have 
${\hat {\vec {\cal T}}}^2 + {\hat {\vec {\cal S}}}^2 = 1$. There is another bilinear form of operators that commutes with all the generators, namely $\hat {\vec {\cal T}} \cdot \hat {\vec {\cal S}}= 0$.

\subsection{Faddeev-Jackiw theory  and path integral for SO(4) fields\label{FJSO4}}

In order to  develop a path integral representation for quantum system whose operators satisfy general Lie algebras, the coherent-state approach \cite{perelomov86} is often used. Here we will instead use the formalism introduced by Faddeev and Jackiw (FJ) \cite{Faddeev88,Govaerts90,Barcelo92}, that provides a way to obtain a classical theory consistent with the algebra of the quantum problem without need of dealing with the structure of the corresponding group. In this approach, no formal distinction is made between different forms of constraints like in Dirac's theory \cite{Dirac64}, where primary and secondary, first class and second class constraints appear. As shown by Faddeev and Jackiw \cite{Faddeev88,Govaerts90,Barcelo92} constraints can be incorporated iteratively
(see also this section and Sec.\ \ref{FJBondFields}), until the basic brackets for the fields can be determined.
Once the classical field-theory is obtained, quantization can proceed via a path integral or in a canonical way. 

Our purpose is to construct a classical Lagrangian first order in time derivatives of the SO(4) fields. The terms containing time derivatives will give the corresponding Berry phases. Following FJ, we call these terms canonical. Since the FJ-formalism is not widely used, we will discuss it in some detail, such that the presentation is to a large extent self-contained. Given the Hamiltonian $H$, that depends on fields $y_A$, we start by writing a Lagrangian first order in the velocities $\dot{y}_A$,
\bea \label{Ly0}
{\mathcal{L}}(y_A, \dot{y}_A)
&=&\sum_{A}\dot{y}_{A}\;K_{A}(y_{A})-H(y_A) \; .
\eea
Then, the associated Euler-Lagrange equations of motion are (summation over repeated indices is assumed)
 \begin{eqnarray}\label{EuLa}
\left[ \frac{\partial{K}_B}{\partial{y_A}}-
\frac{\partial{K_A}} {\partial{y_B}}\right] \,\dot{y}_B-\frac{\partial H}{\partial y_A}=0
\end{eqnarray}
If the matrix defined by
\begin{eqnarray}
M_{AB}\equiv\frac{\partial{K}_B}{\partial{y_A}}-
\frac{\partial{K_A}} {\partial{y_B}}
\end{eqnarray}
is nonsingular,
it is possible to write eq.\ (\ref{EuLa}) as
\begin{eqnarray}\label{doty}
\dot{y}_B =(M_{AB})^{-1} \frac{\partial H}{\partial y_A}
\end{eqnarray}
On the other hand,  from the Hamiltonian formalism the equations of motion are
\begin{eqnarray}\label{dotyH}
\dot{y}_B = \left\{H, y_B \right\}= \frac{\partial H}{\partial y_A} \left\{y_A,y_B \right\} \; ,
\end{eqnarray}
where $\{A,B\}$ denotes the Poisson bracket. 
Then, comparing (\ref{doty}) and (\ref{dotyH}), we see that $(M_{AB})^{-1}$ plays the role of the basic bracket  (or generalized bracket) of the Faddeev-Jackiw theory, i.e.
\begin{eqnarray}
\label{eq:bra}
\left\{y_{A},y_{B}\right\}_{FJ} =\;(M_{AB})^{-1}
\end{eqnarray}
These brackets agree with the Poisson brackets  for an unconstrained theory. The generalized bracket between two quantities $F(y_A)$ and $G(y_A)$ is defined by
\begin{eqnarray}
\label{eq:genbra}
\left\{F,G\right\}_{FJ}=\sum_{AB} \frac{\partial F}{\partial y_A}\left\{y_{A},y_{B}\right\}_{FJ}\frac{\partial G}{\partial y_B}
\end{eqnarray}
From (\ref{eq:bra}) and (\ref{eq:genbra}) it is straightforward to show that the generalized brackets verify all the properties of the usual Poisson brackets.

In our case, we assume that the first-order Lagrangian can be written in terms of the SO(4)-fields as:
\begin{eqnarray}\label{L0TS}
 {\mathcal{L}}(\mathcal{T},\mathcal{S})&=&
{\mathcal{A}}_a^\mathcal{T}\dot{\mathcal{T}^a}+
{\mathcal{A}}_a^\mathcal{S}\dot{\mathcal{S}^a}-V^{(0)}
\; ,
\end{eqnarray}
where $\mathcal{A}_a^{\mathcal{T}}$ and ${\mathcal{A}}_a^{\mathcal{S}}$ are unknown coefficients which must be determined. Since the SO(4)-fields must verify the conditions discussed in Sec.\ \ref{SO4Algebra}, we are in the presence of a constrained theory, where the potential $V^{(0)}$ is
\begin{eqnarray}
V^{(0)}= H+ \xi_1 \Omega_{1}+\xi_{2}\Omega_{2} \; ,
\end{eqnarray}
with $H$ the proper Hamiltonian,
$\xi_1$ and $\xi_2$ Lagrange multipliers, and
\begin{eqnarray}\label{vinculos1}
\Omega_{1} &=&{\vec {\cal{T}}}^{2}+{\vec {\cal{S}}}^{2}- 1 
\; ,
\nonumber\\
\Omega_{2} &=&{\vec {\cal{T}}}\cdot {\vec {\cal{S}}} \; ,
\end{eqnarray}
are  the invariants discussed in Sec.\ \ref{SO4Algebra}, which must be considered here as constraints between the SO(4)-fields.

For the classical Lagrangian (\ref{L0TS}), the set of classical variables in configuration space is
$\{y_{A}\}=\{\mathcal{T}^a,\mathcal{S}^a,\xi_{1},\xi_{2}\}$, the coefficients are
$\{K_A\}=\{{\mathcal{A}}_a^\mathcal{T},{\mathcal{A}}_a^\mathcal{S}\}$, 
and the corresponding equations of motion are
\bea
\label{EqMotionV0}
M_{A B} \, \dot{y}_B = \frac{\partial V^{(0)}}{\partial y_A} \; .
\eea
After constructing the matrix $M_{AB}$, it can be readily seen that it is singular because the coefficients $K_A$ are independent of the variables $\xi_{1},\xi_{2}$. This can be remedied by promoting the constraints into the canonical terms \cite{Barcelo92}. 
For that purpose, we notice first, that when multiplying (\ref{EqMotionV0}) by eigenvectors $v_A^{(i)}$
corresponding to the zero modes, we obtain 
\bea
\label{ZeroModes}
v_A^{(i)} \frac{\partial V^{(0)}}{\partial y_A} = \frac{\partial V^{(0)}}{\partial \xi_i}= 0 \; ,
\eea
putting in evidence, that the zero modes of $M_{A B}$ encode the information of the constraints. 
For consistency the time evolution of the constraints should also obey
\begin{eqnarray}
{\dot \Omega}_i = 0 \; , \quad i,j = 1,2 \; .
\end{eqnarray}
These conditions can be incorporated into the Lagrangian with Lagrange multipliers $\lambda_{1,2}$. 
In this way, discarding total time derivatives, the first iterated Lagrangian results 
\cite{Faddeev88, Barcelo92}
\begin{eqnarray}\label{L1TS}
 {\mathcal{L}}^{(1)}(\mathcal{T},\mathcal{S})&=&
{\mathcal{A}}_a^\mathcal{T}\dot{\mathcal{T}^a}+
{\mathcal{A}}_a^\mathcal{S}\dot{\mathcal{S}^a}+\dot{\lambda_{1}}\,\Omega_1
+\dot{\lambda_{2}}\,\Omega_2-V^{(1)}
\end{eqnarray}
where $V^{(1)}=V^{(0)}|_{\Omega_i=0}=H$. The variables $\xi_{1},\xi_{2}$ have disappeared because the constraints can now be imposed on the canonical part.

The new set of variables is $y_{A}=\{\mathcal{T}^a,\mathcal{S}^a,\lambda_{1}, \lambda_{2}\}$
and the  new  $8\times8$ matrix $M_{AB}$ can be written as
\bea 
M_{AB} & =
& \left(
\begin{array}{rrr}
B^{\cal T} & {\tilde M} & {\tilde F}_1 \\
-{\tilde M}^T & B^{\cal S} & {\tilde F}_2 \\
-{\tilde F}_1^T & -{\tilde F}_2^T & 0
\end{array}
\right) \; , 
\eea
where $B^{\cal T}$ given by
\bea 
B^{\cal T}_{a b} = \frac{\partial {\cal A}^{\cal T}_b}{\partial {\cal T}_a} 
- \frac{\partial {\cal A}^{\cal T}_a}{\partial {\cal T}_b} = \varepsilon_{abc} 
\left({\vec \nabla}_{\cal T} \times {\vec {\cal A}}^{\cal T} \right)_c
\; , 
\eea
and $B^{\cal S}$ given by
\bea 
B^{\cal S}_{a b} = \frac{\partial {\cal A}^{\cal S}_b}{\partial {\cal S}_a} 
- \frac{\partial {\cal A}^{\cal S}_a}{\partial {\cal S}_b} 
= \varepsilon_{abc}\; \left( {\vec \nabla}_{\cal S} \times {\vec {\cal A}}^{\cal S} \right)_c \; ,
\eea
are $3\times3$ antisymmetric matrices. Furthermore, $\tilde M$ is a $3\times3$ matrix given by 
\bea 
\label{DefTildeM}
{\tilde M}_{a b} = \frac{\partial {\cal A}^{\cal S}_b}{\partial
{\cal T}_a} - \frac{\partial {\cal A}^{\cal T}_a}{\partial {\cal
S}_b} \; ,
\eea 
${\tilde F}_1$ is a $2\times3$ matrix given by 
\bea 
{\tilde F}_1 = \left(
\begin{array}{cc}
2 {\cal T}_1 & {\cal S}_1 \\
2 {\cal T}_2 & {\cal S}_2 \\
2 {\cal T}_3 & {\cal S}_3
\end{array}
\right) \; , 
\eea 
and ${\tilde F}_2$ is as ${\tilde F}_1$ with $\vec {\cal T}$ and $\vec {\cal S}$ interchanged.

The unknown coefficients
${\mathcal{A}}_a^\mathcal{T}={\mathcal{A}}_a^\mathcal{T}(\vec
{\cal T} ,\vec {\cal S})$ and
${\mathcal{A}}_a^\mathcal{S}={\mathcal{A}}_a^\mathcal{S}(\vec {\cal T} ,\vec {\cal S})$  must be determined in such a way that the matrix $M_{ab}$ results nonsingular and the SO(4) fields verify the following relations
\bea 
\label{classicalCommutatorTS}
\left\{ {{\cal T}}^a , {{\cal T}}^b \right\}_{FJ} & = &  -i \left[{\hat {\cal T}}^a , {\hat {\cal T}}^b \right]
\; ,
\nonumber\\
\left\{ {{\cal T}}^a ,  {{\cal S}}^b \right\}_{FJ} & = & -i \left[{\hat {\cal T}}^a , {\hat {\cal S}}^b \right] 
\; ,
\nonumber\\
\left\{ {{\cal S}}^a ,  {{\cal S}}^b \right\}_{FJ} & = & -i \left[{\hat {\cal S}}^a , {\hat {\cal S}}^b \right]
\; . 
\eea
These are the classical version of the commutation relations (\ref{CommutatorTS}). From (\ref{eq:bra}) and (\ref{classicalCommutatorTS}) we have
\begin{eqnarray}\label{eq:bra1p}
(M_{AB})^{-1} =\left(
\begin{array}{cccc}
\left\{ {{\cal T}}^a ,  {{\cal T}}^b \right\}_{FJ}& \left\{ {{\cal T}}^a ,  {{\cal S}}^b \right\}_{FJ}
& \left\{{\cal T}_{a},\lambda_{1}\right\}_{FJ}& \left\{{\cal T}_{a},\lambda_{2}\right\}_{FJ}\\\\
 \left\{ {{\cal S}}^a ,  {{\cal T}}^b \right\}_{FJ} & \left\{ {{\cal S}}^a ,  {{\cal S}}^b \right\}_{FJ}
 &  \left\{{\cal S}_{a},\lambda_{1}\right\}_{FJ} &\left\{{\cal S}_{a},\lambda_{2}\right\}_{FJ}\\\\
 -\left\{{\cal T}_{b},\lambda_{1}\right\}_{FJ}&-\left\{{\cal
 S}_{b},\lambda_{1}\right\}_{FJ}&0&\left\{\lambda_{1},\lambda_{2}\right\}_{FJ}\\\\
 -\left\{{\cal T}_{b},\lambda_{1}\right\}_{FJ}&-\left\{{\cal S}_{b},\lambda_{1}\right\}_{FJ}& -\left\{\lambda_{1},\lambda_{2}\right\}_{FJ}& 0\\
\end{array}
\right) \; .
\end{eqnarray}
where in  (\ref{eq:bra1p}) the elements $\left\{{\cal T}_{a},\lambda_{i}\right\}_{FJ}$, $\left\{{\cal S}_{a},\lambda_{i}\right\}_{FJ}$ and
$\left\{\lambda_{1},\lambda_{2}\right\}_{FJ}$ are unknown but unnecessary.
For the determinant of $M_{AB}$, we obtain :
\bea
{\rm det} M_{AB} = 4 D^2 
\;  ,
\eea
where $D = D_{1}+D_{2}$
with
\bea
\label{D1}
D_1 & = &\left[ {\tilde M}_{ab }
\left( {\vec \nabla}_{\cal S} \times {\vec {\cal A}}^{\cal S} \right)_b
- \left( {\vec \nabla}_{\cal T} \times {\vec {\cal A}}^{\cal T} \right)_b
{\tilde M}_{ba} \right]
\left( \vec {\cal S} \times \vec {\cal T} \right)_{a}
\; , \eea
and
\bea \label{D2} D_2 & = & \left[\left( {\vec \nabla}_{\cal S} \times {\vec {\cal A}}^{\cal S} \right) \cdot
\vec {\cal S} \right] \left[\left( {\vec \nabla}_{\cal T} \times {\vec {\cal A}}^{\cal T} \right) \cdot \vec {\cal S} \right]
\nonumber \\ & & 
- \left[\left( {\vec \nabla}_{\cal S} \times {\vec {\cal A}}^{\cal S} \right) \cdot \vec {\cal T} \right]
\left[\left( {\vec \nabla}_{\cal T} \times {\vec {\cal A}}^{\cal T} \right) \cdot \vec {\cal T} \right] 
%\nonumber \\ & & 
+ \frac{1}{2} \left( {\cal T}_a {\cal T}_d - {\cal S}_a {\cal S}_d \right) \varepsilon_{abc} 
\varepsilon_{d e f} {\tilde M}_{e b} {\tilde M}_{f c} \; . 
\eea

Imposing the identities (\ref{classicalCommutatorTS}) in (\ref{eq:bra1p}), and after a long but straightforward algebra it is possible to show that the coefficients must satisfy the following set of equations:
\bea \label {ecu_finales1}
({\vec \nabla}_{\cal T}\times {\mathcal{A}}^{\cal T}+{\vec \nabla}_{\cal S}\times {\mathcal{A}}^{\cal S}) %\;
\cdot \vec{\mathcal{T}}
+({\vec \nabla}_{\cal T}\times {\mathcal{A}}^{\cal S}+{\vec \nabla}_{\cal S}\times {\mathcal{A}}^{\cal T}) %\;
\cdot\vec{\mathcal{S}} & = & -2 \; ,
\nonumber\\
({\vec \nabla}_{\cal T}\times {\mathcal{A}}^{\cal T}+{\vec \nabla}_{\cal S}\times {\mathcal{A}}^{\cal S}) %\;
\cdot \vec {\cal S}
+({\vec \nabla}_{\cal T}\times {\mathcal{A}}^{\cal S}+{\vec \nabla}_{\cal S}\times {\mathcal{A}}^{\cal T}) %\;
\cdot \vec {\mathcal{T}} & = & 0 \; ,
\nonumber \\
\left( {\vec \nabla}_{\cal S} \times {\vec {\cal A}}^{\cal S} 
- {\vec \nabla}_{\cal T} \times {\vec {\cal A}}^{\cal T} \right) \cdot 
\left(\vec {\cal T} \times \vec {\cal S} \right) 
%\nonumber \\ 
+ \left({\tilde M}_{a b} + {\tilde M}_{b a} \right) 
({\cal T}_a {\cal T}_b - {\cal S}_a {\cal S}_b) & &
\nonumber \\ 
+ {\tilde M}_{a a} \left( {\vec {\cal S}}^2-{\vec {\cal T}}^2 \right)
+ \sum_a {\tilde M}_{a a} \left( {\cal S}_a^2-{\cal T}_a^2 \right)
& = & 0 \; ,
\nonumber \\
D  & = & - 1 \; .
\eea
A possible solution compatible with the equations above is
\bea \label {ecu_finales2}
{\vec \nabla}_{\cal T} \times {\mathcal{A}}^{\cal T}  =  {\vec \nabla}_{\cal S}\times {\mathcal{A}}^{\cal S} \; ,
\quad
{\vec \nabla}_{\cal S} \times {\mathcal{A}}^{\cal T}  =  {\vec \nabla}_{\cal T}\times {\mathcal{A}}^{\cal S} \; ,
\quad
\tilde{M}_{ab}  =  -\tilde{M}_{ba} 
 \; ,
\eea
with
\bea
\label{Solution1}
{\vec \nabla}_{\cal T} \times {\vec {\cal A}}^{\cal T}  =  - \vec {\cal T} \; ,
\quad
{\vec \nabla}_{\cal S} \times {\vec {\cal A}}^{\cal T} =  - \vec {\cal S} \; , 
\quad
{\tilde M}_{a
b}  = - \varepsilon_{abc} {\cal S}_c \; ,
\eea
as can be easily verified.
We notice furthermore that, from the last two equations in (\ref{Solution1}) 
we have
\bea 
\label{derivCruzadas}
 \frac{\partial {\cal A}^{\cal S}_b}{\partial {\cal T}_a}
= \frac{\partial {\cal A}^{\cal T}_a}{\partial {\cal S}_b} \; .
\eea
The solutions displayed above are, however, not the most general ones. In the Appendix we show explicit forms of ${\vec {\cal A}}^{\cal T}$ and  ${\vec {\cal A}}^{\cal S}$, that obey eqs.\ (\ref{ecu_finales1}) but not
all eqs.\  (\ref{Solution1}). Nevertheless, it is also shown in the Appendix, that both forms lead to the same Berry phase in the gradient expansion.

Having obtained an effective Lagrangian with 
%the appropriate 
a set of coefficients
${\mathcal{A}}_{\alpha}^\mathcal{T}$, ${\mathcal{A}}_{\alpha}^\mathcal{S}$ and constraints (\ref{vinculos1})
we can write the path integral for the partition function as usual \cite{Gitman90}
\bea 
\label{Z_STau} 
\mathcal{Z }& = & \int {\cal D} \vec {\cal T}
\, {\cal D} \vec {\cal S} \, \delta \left( \vec {\cal T} \cdot
\vec {\cal S} \right) \, \delta \left({\vec {\cal T}}^2 + {\vec
{\cal S}}^2 - 1\right) \, {\rm e}^{-S} \; , 
\eea 
with the Euclidian action
\bea
\label{ActionTauS} 
S & = & \int {\rm d} \tau \left\{ -i \sum_j
\left( {\vec {\cal A}}^{\cal T}_j \cdot \partial_\tau {\vec {\cal T}}_j +
{\vec {\cal A}}^{\cal S}_j \cdot \partial_\tau {\vec {\cal S}}_j \right)
+ H \left[\vec {\cal T}, \vec {\cal S} \right] \right\} \; .
 \eea
Here we would like to remark a few features. In the measure of the path integral (\ref{Z_STau})
only the two constraints enforced by $\delta$-functions appear. No other functional between fields is present because the determinant of the matrix $M_{AB}$ is a constant ($D=-1$).
%last expression in eq.\ 
%(\ref {ecu_finales2})
The first term in the action (\ref{ActionTauS}), i.e.\ the Berry phase, depends on the coefficients 
${\vec {\cal A}}^\mathcal{T}$ and ${\vec {\cal A}}^\mathcal{S}$, that have to fulfill (\ref{ecu_finales1}). Hence, they are defined up to a gauge transformation, 
\bea
{\cal A}_a^{\cal T} \rightarrow {\cal A}_a^{\cal T} + \frac{\partial \Lambda}{\partial {\cal T}_a} \; , \quad
{\cal A}_a^{\cal S} \rightarrow {\cal A}_a^{\cal S} + \frac{\partial \Lambda}{\partial {\cal S}_a} \; , 
\eea
with $\Lambda = \Lambda ({\vec {\cal A}}^\mathcal{T}, {\vec {\cal A}}^\mathcal{S})$ a scalar function,
in a similar way as in the SU(2) case.  
In this way, we have obtained a path integral for the $SO(4)$ fields that
is fully consistent with the algebra of the quantum problem.

\section{Bond-fields  formulation of bond variables}\label{bondF}
\subsection{Bond-fields for spin-states on a bond}\label{bondF1}

It is  usual  for constrained systems, where it is necessary to deal with operators that do not satisfy canonical commutation rules, to introduce slave particles by decoupling the original operators \cite{auerbach94}. In the frame of the Hubbard operators $X^{\mu \nu}$ defined in (\ref{OpHubb}),
the following decoupling can be used,
\begin{eqnarray}
\label{Xop_st}
X^{a b}=t^{\dagger}_a t^{}_b \; , \qquad \qquad X^{0 a}= s^{\dagger} t^{}_a \; ,
\end{eqnarray}
with $a, b = 1,2,3$, and where $\{t^{}_a,\;t^{\dagger}_a,\;s,\;s^{\dagger}\}$ satisfy bosonic commutation rules
\begin{eqnarray}
\label{eq:brabf}
\left[t^{}_a , t^{\dagger}_b\right]=\delta_{a b} \; ,
\qquad \qquad 
\left[s^{},s^{\dagger}\right]=1 \; ,
\qquad \qquad
\left[s^{\dagger},t_a \right]=0 \; .
\end{eqnarray}
Using (\ref{Xop_st}) it is easy to show that the commutation rules (\ref{CommutatorX}) are satisfied. In addition, the completeness condition (\ref{CompletnessX}) can be written as 
\begin{equation}
\label{Completness_st}
s^{\dagger}\;s+t^{\dagger}_a t^{}_a= 1 \; .
\end{equation}

From eqs.\ (\ref{ST_operators}), (\ref{SpinFromTS}) and (\ref{Xop_st}) we obtain
\begin{eqnarray}\label{Sst}
S_{(1)}^a & = & 
\frac{1}{2} \left(s^{\dagger} t^{}_a+t^{\dagger}_a s-i 
\varepsilon_{a b c}\;t^{\dagger}_b t^{}_c \right) \; ,
\nonumber\\
S_{(2)}^a & = & 
\frac{1}{2} \left(-s^{\dagger} t_a - t^{\dagger}_a s
- i \varepsilon_{a b c} t^{\dagger}_b t^{}_c \right) \; ,
\end{eqnarray}
that exactly correspond to the relation between spin- and bond-operators introduced by Sachdev and Bhatt \cite{sachdev90}.

As the path integral (\ref{Z_STau}) shows, 
%it is evident that 
only four real variables are independent. Therefore, on passing from bond-operators to bond-fields, four constraints 
are needed. One of them is the completeness condition
\begin{eqnarray}
\label{newvinvarphi1}
\varphi_{1} & \equiv & s^*  s + t^*_a t^{}_a -1 = 0
\; .
\end{eqnarray}
In order to obtain the remaining conditions, we use
the well known CP$^1$ representation \cite{auerbach94} for the spin-fields
${\vec S}_{(1)}$ and ${\vec S}_{(2)}$,
\begin{eqnarray}
\label{SP0}
  S_{(1)}^a= \frac{1}{2} \, \bar z \sigma^a z \; ,
\qquad \qquad  \qquad
  S_{(2)}^a= \frac{1}{2} \,\bar \omega \sigma^a \omega \; ,
\end{eqnarray}
where $\sigma^a$ are the Pauli matrices. The CP$^1$ fields, $\bar z=(z^*_{1},z^*_{2})$ and 
$\bar \omega=(\omega^*_{1},\omega^*_{2})$, fulfill the conditions
\begin{eqnarray}
\label{zwP}
  \bar z z= 1 \; , \qquad \qquad  \qquad
  \bar \omega \omega= 1 \; .
\end{eqnarray}
On the basis of eqs.\ (\ref{Sst}) and (\ref{SP0}), the following relations between bond-fields and CP$^1$ variables can be obtained:
\begin{eqnarray}
\label{Trans1}
  s & = & \frac{1}{\sqrt{2}}\; \left({z}^{}_{1}\;\omega^{}_{2}\;-\;
  z^{}_{2}\;\omega_{1} \right) \; ,
\nonumber \\
t_{1} & = & \frac{1}{\sqrt{2}}\;\left({z}_{2}\;\omega_{2}\;-\; z_{1}\;\omega_{1} \right) \; ,
\nonumber \\
  t_{2} & = & -\frac{i}{\sqrt{2}}\; \left({z}_{1}\;\omega_{1}\;+\;
  z_{2}\;\omega_{2} \right) \; ,
\nonumber \\
  t_{3} & = & \frac{1}{\sqrt{2}}\;\left({z}_{1}\;\omega_{2}\;+\;
  z_{2}\;\omega_{1} \right) \; .
 \end{eqnarray}
These equations lead to two additional constraints for the bond-fields:
\begin{eqnarray}
\label{newvin2}
\varphi_{2} & \equiv & s s-t_a t^{}_a = 0 \; ,
\nonumber\\
\varphi_{3} & \equiv & s^* s^* - t^*_a t^*_a = 0
\; .
\end{eqnarray}
As can be seen from (\ref{Sst}), 
the theory defined in terms
of the bond-fields contains a gauge degree of freedom such that, the
remaining constraint will appear as a gauge-fixing condition, that we discuss in the next section.

\subsection{Faddeev-Jackiw theory and path integral formulation for bond-operators\label{FJBondFields}}
In this section we will develop the FJ-formalism for bond-operators. The procedure will parallel that of Sec.\  \ref{FJSO4}, but due to a gauge freedom, it will be iterated, until the condition for gauge fixing is incorporated in the theory.
 
Our starting point is the classical Lagrangian
\begin{equation}
\label{Ls1}
  {\mathcal{L}} = -\frac{i}{2}(\dot{s^*}s-s^*\dot{s}+\dot{t}^*_{a} t^{}_{a}-t^*_{a} \dot{t}^{}_{a})-
 V^{(0)} \; .
\end{equation}
where $ V^{(0)}=H(s^*,s,t^*_{a}, t^{}_{a}) + \xi_i \varphi_i\;$, $H(s^*,s,t^*_{a}, t^{}_{a})$ is the spin Hamiltonian written in terms of the bond-fields, $ \xi_i $  are 
Lagrange multipiers and
the constraints $\varphi_i$ with $i=1,2,3$ were defined in eqs.\ 
(\ref{newvinvarphi1}) and (\ref{newvin2}) 
in the last subsection. As in previous bond-operator treatments \cite{sachdev90}, we have adopted the usual kinetic Lagrangian for bosons. As we will see below, this is consistent with the $SO(4)$ algebra.

From the Lagrangian (\ref{Ls1}), the set of classical variables  is
$\{y_{A}\}=\{q_{\alpha}, \xi_{i}\}$, where $q_{\alpha} = \left\{s,s^*,t^{}_{a},t^*_{a} \right\}$, with the index
$\alpha=1,\dots,8$.
The corresponding matrix $M_{AB}$ is singular, because the coefficients
$K_A$
do not contain the variables $\xi_{i}$. In this case the matrix $M_{AB}$ has three zero eigenvectors $\vec v^{(i)}$ . Multiplying
the Euler-Lagrange equations 
(\ref{EqMotionV0})
by $\vec v^{(i)}$, 
we obtain as in Sec.\ \ref{FJSO4}
\begin{eqnarray}
\label{newconstst}
 v^{(i)}_{A}\;\frac{\partial V^{(0)}}{\partial y_{A}}=\frac{\partial V^{(0)}}
 {\partial \xi_i} 
 =0 \; ,
\end{eqnarray}
that leads to $\varphi_i = 0$.
As  described in Sec.\ \ref{FJSO4} and in Refs.\ \cite{Faddeev88, Barcelo92} we incorporate the constraints (\ref{newconstst}) into the kinetic part of the Lagrangian using new Lagrange multipliers
$\lambda_{i}$. The first iterated Lagrangian results
\begin{eqnarray}
\label{L1BondF}
 {\mathcal{L}}^{(1)}(\mathcal{T},\mathcal{S})&=&
-\frac{i}{2} \left(\dot{s}^* s-\dot{s} s^*+\dot{t}^*_{a} t^{}_{a} -\dot{t}^{}_{a} t^*_{a} \right)+\dot{\lambda}_{i} \,\varphi_i - V^{(1)} \; ,
\end{eqnarray}
where $V^{(1)}=V^{(0)}|_{\varphi_i=0}=H$.
For the Lagrangian (\ref{L1BondF}) the new set of variables is
$y_{A}=\{q_{\alpha},\lambda_{i}\}$ and the  new  matrix $M_{AB}$
is
\bea M_{AB} & = & \left[
\begin{array}{crc}
f_{\alpha\beta} &  & \frac{\partial \varphi_{i}}{\partial q_{\alpha}} \\
-\left(\frac{\partial \varphi_{i}}{\partial q_{\beta}}\right)^T &&0 \\
\end{array}
\right] \; ,
\eea
where
\bea f_{\alpha\beta}  & = & \left(
\begin{array}{rrrr}
0 & i & 0&0 \\
-i &0&0 &0\\
0 &0&0 &i\,\delta_{ab}\\
0 &0&-i\,\delta_{ab}&0 \\
\end{array}
\right) \; .
\eea
Note that $f^{-1}_{\alpha\beta}$ is the matrix formed by the classical extension of the commutations relations for bond fields.

As the matrix $M_{AB}$ is antisymmetric with an odd number of rows and columns $(11\times11)$, 
it is singular. After the first iteration, $M_{AB}$ has only one zero-eigenvector $\vec u$ \cite{montani93}
\bea 
\label{uautovector} 
\vec u^{T}&=& \left(-\frac{\partial
\varphi_{1}}{\partial
q_{\alpha}}\;f^{-1}_{\alpha\beta},1,0,0 \right)
= \left(-i\,s,i\,s^*,-i\,t^{}_{a},i\,t^*_{a},1,0,0 \right)\;.
\eea 
As before, the new constraint is obtained  multiplying  the Euler-Lagrange equation 
(\ref{EqMotionV0})
by the zero-eigenvector $\vec u$, i.e.\ the new constraint is
\bea
\label{FJconsistent} 
0&=& u_{A}^{T}  \frac{\partial
V^{(1)}}{\partial y_{A}}=-\frac{\partial \varphi_{1}}{\partial
q_{\alpha}}\;f^{-1}_{\alpha\beta}\, \frac{\partial V^{(1)}}{\partial
q_{\beta}}=\left\lbrace \varphi_1, V^{(1) }\right\rbrace
=\dot{\varphi_1} \; ,
\eea
where we have used the equations (\ref{uautovector}) and (\ref{eq:genbra}). Therefore, no new constraint  arises from  equation (\ref{FJconsistent}) because $\varphi_{1}$ is already a constraint and its time evolution is also a constraint for consistency. This fact is related to the existence of a gauge degree of
freedom  associated with the completeness condition $\varphi_{1}$.

In order to obtain a gauge fixing condition $\varphi_4$ as a new constraint, we will repeat the process introducing a new Lagrangian ${\mathcal{L}}^{(2)}$ 
\begin{eqnarray}
\label{L2BondF}
 {\mathcal{L}}^{(2)}(\mathcal{T},\mathcal{S})&=&
-\frac{i}{2} \left(\dot{s}^* s- \dot{s} s^* + \dot{t}^*_{a} t^{}_{a} -\dot{t}^{}_{a} t^*_{a} \right)
+\dot{\lambda}_{i} \varphi_i +\dot{\lambda}_{4} \varphi_4 -V^{(2)} \; ,
\end{eqnarray}
where $V^{(2)}=V^{(1)}|_{\varphi_i=0}=H$ and the set of variables of the new configuration space is
$\{y_{A}\}=\{q_{\alpha},\lambda_{i},\lambda_{4}\}$.
We will choose a gauge fixing condition $\varphi_4$ in such a way that the new matrix  $M_{AB}$ is nonsingular. In other words
\begin{eqnarray}
\label{detst}
det\;\left[ M_{AB}\right]& = &16\;\left[ \frac{\partial \varphi_{4}}{\partial s}\,s-\frac{\partial \varphi_{4}}{\partial s^{*}}\,s^{*}+\frac{\partial \varphi_{4}}{\partial t_{a}}\,t_{a}
-\frac{\partial \varphi_{4}}{\partial t_{a}^{*}}\,t_{a}^{*}\right] ^2
= 16\;\Lambda^{2}
\end{eqnarray}
must be different from zero.
Computing the inverse of $M_{AB}$, we obtain the following FJ
brackets between bond fields:
\begin{eqnarray}\label{FinalFJBst}
\{s,s\}_{FJ} & = & \{s^*,s^*\}_{FJ} = 0 \; ,
\nonumber\\
\{s,s^*\}_{FJ} & = & i \left(1-s s^*\right)+ 
\left(s\;w_{1}^* -s^* w_{1} \right) \; ,
\nonumber\\
\{s,t_{a}\}_{FJ} & = & t_{a} w_{1}- s v_{a} \; ,
\nonumber\\
\{s,t^*_{a}\}_{FJ} & = & i s^* t^{}_{a} + \left( s v^*_{a}-t^*_{a} {w}_{1} \right) \; ,
\nonumber\\
\{s^*,t_{a}\}_{FJ} & = & -i s t_{a}^* 
+ \left(s^* {v}_{a}-t^{}_{a} {w}^*_{1} \right) \; ,
\nonumber\\
\{s^*,t^*_{a}\}_{FJ} & = & t^*_{a} {w}^*_{1} - s^* v^*_{a} \; ,
\nonumber\\
\{t_{a},t_{b}\}_{FJ} & = & -\left( t_{a} {v}_{b}-t_{b} {v}_{a} \right) \; ,
\nonumber\\
\{t^{}_{a},t^*_{b}\}_{FJ} & = & i \left( \delta_{ab} - t_{a}^* t^{}_{b}\right)
+\left( t^{}_{a} v^*_{b} -t^*_{b} {v}_{a} \right) \; ,
\nonumber\\
\{t^*_{a},t^*_{b}\}_{FJ} & = & - \left(t^*_{a} v^*_{b}-t^*_{b} v^*_{a} \right) \; ,
\end{eqnarray}
where
\bea
w_{1} & = & -\frac{i}{\;\Lambda}\left[ (1-s^* s)
\frac{\partial \varphi_{4}}{\partial s^*} + s^* 
\frac{\partial \varphi_{4}}{\partial t_{a}^*}\,t_{a}\right]\;, 
\nonumber\\
v_{a} & = & -\frac{i}{\;\Lambda}\left[
(\,\delta_{ab}-t_{a}^* t_{b})
\frac{\partial \varphi_{4}}{\partial t_{b}^*}
+\frac{\partial \varphi_{4}}{\partial s^*} s t_{a}^*\right]
\;,
\eea
and $a=1,2,3$. Note that on taking into account the constraints, the FJ brackets are different
from the usual bosonic commutation rules. As expected, using (\ref{FinalFJBst}), for any explicit form of a gauge fixing, the SO(4) algebra is fulfilled.

Finally, the partition function for the bond-fields can be written as \cite{Gitman90}
\begin{eqnarray}
\label{PIst0}
{\mathcal{Z}} = \int\,{\mathcal{D}}s^{*}\,{\mathcal{D}}s\,{\mathcal{D}}t^{*}_{a}\,{\mathcal{D}}t_{a} \,
(det\;\left[M_{AB}\right])^{1/2} \,\delta[\varphi_{4}] \; \delta[\varphi_{1}] \, \delta[\varphi_{2}] \,
\delta[\varphi_{3}] \; 
e^{-S
%i\int\;L(s^{*},\,s,\,t^{*}_{a},\,t_{a}) \,dt
}
\; ,
 \end{eqnarray}
where 
\begin{equation}
\label{ActionBondFiledsGeneral}
S = \int {\rm d} \tau \, \left[
(s^* \dot{s} + t^*_a \dot{t}^{}_a ) + H 
\right] \; ,
\end{equation}
and
$(det\;\left[ M_{AB}\right])^{1/2} =4\;\Lambda$ is equivalent to the Faddeev-Popov (FP) determinant $\Delta_{FP}$ in gauge theories \cite{Gitman90}.

Before closing this section, we would like to remark that bond-operators were frequently used at the mean field level, or closely related approximations,  where the full form of (\ref{PIst0}) is not respected \cite{sachdev90,chubukov91,kotov98,matsumoto02,vojta99}.
There,
only the effective Lagrangian $L$ and $\varphi_1$ are consider 
while
$\varphi_2$, $\varphi_3$, $\varphi_4$ and
$\Delta_{FP}=(det\;\left[ M_{AB}\right])^{1/2}$ 
are missing. 
 While on a mean-field level the measure and constraints are not crucial, their presence is important when considering the effect of fluctuations. We show in Sec.\ \ref{Constraints and measure} how, dealing with them, leads to the correct long-wavelength effective action.

We will show in the following with the one-dimensional spin-Peierls systems as an example,
how the corresponding continuum theory can be recovered starting with a path integral for the dimerized state.

\section{Spin-Peierls chain\label{SPchain}}
As an application of the formulations developed above, we consider a spin-Peierls chain. On the one hand, this is  a well known system. On the other hand, the corresponding field-theory has a topological term, closely related to the one present in the field-theory for the antiferromagnetic Heisenberg model. In order to obtain it, a proper treatment of the fluctuations is needed. 
    
The spin-Peierls Hamiltonian in one dimension can be written as 
\be
\label{hamil} 
H_{SP} = J \sum_i \left[ 1 + (-1)^i \Delta \right]
{\vec S}_i \cdot {\vec S}_{i+1} \; , 
\ee 
where $\Delta < 1$ indicates the degree of dimerization. For $\Delta =1$, the system breaks down into a set of decoupled dimers, while for $\Delta = 0$, it reduces to the antiferromagnetic Heisenberg chain.

We can introduce strong bonds given by $(i,i+1)$ with $i=2j$, $j \in \mathbb{Z}$. Using this notation the spin-Peierls Hamiltonian (\ref{hamil}) reads
\bea 
H_{SP} & = & J \sum_j 
\left[ \left( 1 + \Delta \right) {\vec S}_{j,(1)} \cdot {\vec S}_{j,(2)} 
+ \left( 1 - \Delta \right) {\vec S}_{j,(2)} \cdot {\vec S}_{j+1,(1)} \right] \; ,
\eea
where the dimerization pattern is described by bonds labeled by the index $j$. 

\subsection{Continuum limit with SO(4) fields}\label{SO(4)formulation1}
Close to the point where the system goes over to the state appropriate for the antiferromagnetic Heisenberg chain, a large correlation length should be expected, such that the continuum limit is appropriate. 
Using (\ref{SpinFromTS}), $H_{SP}$ can be written in terms of the SO(4)-fields as follows
\bea 
\label{HamiltonianSPWithSO4} 
H_{SP} & = & \frac{J}{4}  \sum_j \Bigg[ \left(1 + \Delta \right) 
\left( {\vec {\cal T}}^2_j - {\vec {\cal S}}^2_j \right) 
+ \left( 1 - \Delta \right) \left({\vec {\cal T}}_j - {\vec {\cal S}}_j \right) \cdot 
\left({\vec {\cal T}}_{j+1} + {\vec {\cal S}}_{j+1} \right)
\Bigg] \; . 
\eea

According to (\ref{Z_STau}), the path-integral for this Hamiltonian is given by 
\bea Z & = &
\int {\cal D} \vec {\cal T} \, {\cal D} \vec {\cal S} \, \delta
\left( \vec {\cal T} \cdot \vec {\cal S} \right) \, \delta
\left({\vec {\cal T}}^2 + {\vec {\cal S}}^2 - 1\right) \, {\rm
e}^{-S} \; , 
\eea 
with the action in imaginary time
\bea 
\label{Action1} 
S & = &
\int {\rm d} \tau \left\{ -i \sum_j \left( {\vec {\cal A}}^{\cal T}_j
\cdot \partial_\tau {\vec {\cal T}}_j + {\vec {\cal A}}^{\cal S}_j \cdot
\partial_\tau {\vec {\cal S}}_j \right) + H_{SP} \left[\vec {\cal
T}, \vec {\cal S} \right] \right\} \; . 
\eea
We approach the continuum limit by performing a gradient expansion around a configuration that is appropriate for large $\Delta$, where the Hamiltonian is dominated by 
\bea 
H_{\Delta} & = & \frac{J}{4}
\sum_j  \left( 1 + \Delta \right) \left( {\vec {\cal T}}^2_j -
{\vec {\cal S}}^2_j \right) \; . 
\eea 
From a mean-field point of view, 
the lowest energy is obtained by maximizing ${\vec {\cal S}}^2$ and consequently minimizing ${\vec {\cal T}}^2$. The classical configuration of lowest energy is given by ${\vec {\cal S}}^2 = 1$, so that we take
\bea 
{\vec {\cal S}}_j & = & C_j {\vec n}_j 
\eea 
with ${\vec n}_j^2 = 1$, such that the condition 
${\vec {\cal T}}^2_j + {\vec {\cal S}}^2_j = 1$
leads to
\bea C_j & = & \sqrt{1-{\vec {\cal T}}^2_j} \; . 
\eea 
Since ${\vec {\cal T}}_j$ is proportional to the change of $ {\vec {\cal S}}_j$, i.e.\ 
\bea 
{\vec {\cal T}}_j \sim \partial_\mu {\vec {\cal S}}_j \; , 
\eea 
the fields entering the action are 
\bea
\label{FieldExpansion} {\vec {\cal S}}_j & = & {\vec n}_j
\sqrt{1-a^2 {\vec \ell}_j^2} \; ,
\nonumber \\
{\vec {\cal T}}_j & = & a {\vec \ell}_j \; , 
\eea 
where $a$ is the lattice constant of the new lattice. In addition, ${\vec \ell}\cdot{\vec n}=0$. 

After defining the fields for the gradient expansion, we consider the different pieces of the action. First we have 
\bea
\label{T2MinusS2} 
{\vec {\cal T}}^2_j - {\vec {\cal S}}^2_j & = &
%a^2 {\vec \ell}_j^2 - \left(1-a^2 {\vec \ell}_j^2 \right) = 
2 a^2
{\vec \ell}_j^2 -1 \; . 
\eea 
Hence, the modes described by the field ${\vec \ell}_j$ are massive. Since we are in 1+1 dimensions,
we perform the gradient expansion up to ${\cal O} \left(a^2\right)$. For the rest of the terms coming from the
Hamiltonian we have, 
\bea 
\label{CouplingsTS} 
{\vec {\cal T}}_j \cdot {\vec {\cal T}}_{j+1} 
& \simeq & a^2 {\vec \ell}_j^2 \; ,
\nonumber \\
{\vec {\cal T}}_j \cdot {\vec {\cal S}}_{j+1}  
& \simeq & a^2 {\vec \ell}_j \cdot \partial_x {\vec n}_j \; ,
\nonumber \\
{\vec {\cal S}}_j \cdot {\vec {\cal T}}_{j+1} 
& \simeq & a^2 {\vec n}_j  \cdot \partial_x {\vec \ell}_j \; ,
\nonumber \\
{\vec {\cal S}}_j \cdot {\vec {\cal S}}_{j+1}  
& \simeq &
 1 - a^2 {\vec \ell}_j^2 + \frac{1}{2} a^2  {\vec n}_j  \cdot \partial_x^2  {\vec n}_j \; .
\eea 
Putting all the contributions together, and going over to the continuum, we have 
\bea 
H_{SP} \left[\vec {\cal T}, \vec {\cal S} \right] & \rightarrow & 
J a \int {\rm d} x \,  \left\{ {\vec \ell}^2 + \left( 1 - \Delta \right)
\left[ \frac{1}{2} {\vec \ell} \cdot \partial_x {\vec n} +
\frac{1}{8} \left(\partial_x  {\vec n} \right)^2 \right] \right\}
\; . 
\eea

Next we deal with the Berry phase.  Here we have 
\bea
\label{BerryPhase} 
S_B & = & -i \int {\rm d} \tau \sum_j \left(
{\vec {\cal A}}^{\cal T}_j \cdot \partial_\tau {\vec {\cal T}}_j + {\vec {\cal A}}^{\cal S}_j \cdot 
\partial_\tau {\vec {\cal S}}_j \right) \; .
\eea 
Since from the discussion above, the SO(4)-fields have only a smooth spatial dependence, we can concentrate on one site. Furthermore, since from (\ref{FieldExpansion}) $\mid \vec {\cal T} \mid \sim a$, we can expand the vector potentials accordingly. In lowest order we have from (\ref{Solution1}), 
\bea
\label{AsInZerothOrder} 
\left. {\vec \nabla}_{\cal T} \times {\vec {\cal A}}^{\cal T} \right|_{{\cal T}=0} 
= \left. {\vec \nabla}_{\cal S} \times {\vec {\cal A}}^{\cal S}
\right|_{{\cal T}=0} =  0 \; . 
\eea 
Hence, we can write 
\bea
\label{GaugeAway} 
{\cal A}^{\cal T}_a \left(\vec {\cal T} = 0, \vec
{\cal S} \right) & = & \frac{\partial }{\partial {\cal T}^a}
\phi^{\cal T}   \left(\vec {\cal T} , \vec {\cal S} \right) \; ,
\nonumber \\
{\cal A}^{\cal S}_a \left(\vec {\cal T} = 0, \vec {\cal S} \right) & =
& \frac{\partial }{\partial {\cal S}^a} \phi^{\cal S}   \left(\vec
{\cal T} , \vec {\cal S} \right) \; , 
\eea 
where $\phi^{\cal T} $ and $\phi^{\cal S} $ are scalar functions. Such terms can be gauged away when necessary.

The expansion of the fields ${\vec {\cal A}}^{\cal T}$ and ${\vec {\cal A}}^{\cal S}$ in powers of ${\cal T}^a$ leads to 
\bea 
{\cal A}^{\cal T}_a \left(\vec {\cal T} , \vec {\cal S} \right) & = & 
{\cal A}^{\cal T}_a \left(\vec {\cal T} = 0, \vec {\cal S} \right) 
+ \left. \frac{\partial {\cal A}^{\cal T}_a}{\partial {\cal
T}^b} \right|_{{\cal T}=0}  {\cal T}^b + {\cal O} \left(a^2\right)
\; ,
\nonumber \\
{\cal A}^{\cal S}_a \left(\vec {\cal T} , \vec {\cal S} \right) & = &
{\cal A}^{\cal S}_a \left(\vec {\cal T} = 0, \vec {\cal S} \right) +
\left. \frac{\partial {\cal A}^{\cal S}_a}{\partial {\cal T}^b}
\right|_{{\cal T}=0}  {\cal T}^b + {\cal O} \left(a^2\right) \; ,
\eea 
where due to the presence of one time derivative in the Berry phase, we need to consider only terms up to ${\cal O} (a)$. Introducing the expansion into (\ref{BerryPhase}) leads to 
\bea
\label{BerryPhaseExpansion} 
{\vec {\cal A}}^{\cal T} \cdot \partial_\tau
\vec {\cal T} + {\vec {\cal A}}^{\cal S} \cdot \partial_\tau \vec {\cal S} 
& = & 
{\vec {\cal A}}^{\cal T} \left( 0, \vec {\cal S} \right) \cdot \partial_\tau \vec {\cal T} 
+  \left. \frac{\partial {\cal A}^{\cal T}_a}{\partial {\cal T}^b} \right|_{{\cal T}=0} 
{\cal T}^b \, \partial_\tau {\cal T}^a 
\nonumber \\ & & 
+ {\vec {\cal A}}^{\cal S} \left( 0, \vec {\cal S} \right) \cdot \partial_\tau \vec {\cal S}
+  \left. \frac{\partial {\cal A}^{\cal S}_a}{\partial {\cal T}^b} \right|_{{\cal T}=0}  
{\cal T}^b \, \partial_\tau {\cal S}^a \; .
\eea 
From (\ref{Solution1})  we have \bea
\frac{\partial {\cal A}^{\cal T}_a}{\partial {\cal T}^b} & = &
\frac{\partial {\cal A}^{\cal T}_b}{\partial {\cal T}^a} +
\varepsilon^{abc} {\cal T}^c \; ,
\nonumber \\
\frac{\partial {\cal A}^{\cal S}_a}{\partial {\cal T}^b} & = &
\frac{\partial {\cal A}^{\cal S}_b}{\partial {\cal T}^a} +
\varepsilon^{abc} {\cal S}^c \; . 
\eea 
such that using
(\ref{derivCruzadas}), we arrive at 
\bea 
{\vec {\cal A}}^{\cal T} \cdot \partial_\tau
\vec {\cal T} + {\vec {\cal A}}^{\cal S} \cdot \partial_\tau \vec {\cal S} 
& = &
\partial_\tau \left( {\vec {\cal A}}^{\cal T} \cdot \vec {\cal T} \right)
+ {\vec {\cal A}}^{\cal S} \left( 0, \vec {\cal S} \right) \cdot
\partial_\tau \vec {\cal S} + \varepsilon^{abc} \partial_\tau
{\cal S}^a {\cal T}^b {\cal S}^c \nonumber \\ & = &
\varepsilon^{abc} \partial_\tau {\cal S}^a {\cal T}^b {\cal S}^c
\; , \eea 
where the first term can be discarded since it is  a total time derivative and the second one can be gauged away, as discussed after eq.\ (\ref{GaugeAway}). Inserting the relations (\ref{FieldExpansion}), we finally have for the Berry phase in the continuum limit, 
\bea
\label{BerryPhasep} 
S_B & = & 
-i \int {\rm d} \tau \, {\rm d} x \, \vec \ell \cdot \left(
\vec n \times \partial_\tau \vec n \right) \; . 
\eea

Going back to (\ref{Action1}), we have for the action after the gradient expansion, 
\bea 
\label{FieldAction} 
S & = & -i \int {\rm d} \tau \, {\rm d} x \, \vec \ell \cdot \left( \vec n \times \partial_\tau \vec n \right) 
\nonumber \\ & & 
+J a \int {\rm d}
\tau \,  {\rm d} x \,  \left\{ {\vec \ell}^2 + \left( 1 - \Delta
\right) \left[ \frac{1}{2} {\vec \ell} \cdot \partial_x {\vec n} +
\frac{1}{8} \left(\partial_x  {\vec n} \right)^2 \right] \right\}
\; . 
\eea
At this point $\vec \ell$ can be integrated out. Since the action is quadratic in this field, we can simply consider the saddle point for $\vec \ell$, that leads to
\bea \ell^a & = & 
\frac{1}{2 Ja} \left[ i \varepsilon^{abc} n^b \partial_\tau n^c 
- \frac{Ja}{2} \left( 1 - \Delta \right) \partial_x n^a
\right] \; . 
\eea
Inserting this into the action, we finally obtain
\bea 
\label{actioninnl}
S & = & \int {\rm d} \tau \, {\rm d} x \, \left[
\frac{1}{4Ja} \left( \partial_\tau \vec n \right)^2
+ \frac{J a}{16} \left(1-\Delta^2 \right) \left(\partial_x  {\vec
n} \right)^2
\right] 
\nonumber \\ & & \quad
+ i \, \frac{(1-\Delta)}{4} \int {\rm d}
\tau \, {\rm d} x \, \vec n \cdot \left(\partial_\tau \vec n
\times \partial_x \vec n \right) \; . 
\eea
The coupling constant and spin-wave velocity are given by
\bea
g&=&\frac{4}{(1-\Delta^{2})^{1/2}} \; ,
\eea
and 
\bea
c & = & \frac{J}{2} \left[1-\Delta^{2} \right]^{1/2} \; .
\eea
The same result can be achieved performing the same treatment as for the Heisenberg model \cite{auerbach94}.

 The action (\ref{actioninnl}) corresponds to an O(3) non-linear $\sigma$-model with a topological term
 $i \theta Q$, with $\theta = S (1-\Delta)/2$, 
for $S=1/2$, and $Q$ given by the integral in the last term of (\ref{actioninnl}), acquiring values $4 n \pi$,
 with $n$ an integer. For the antiferromagnetic Heisenberg chain ($\Delta = 0$), it is responsible for the absence
 of a gap for half-integer $S$, while a gap is present for $S$ integer, as first discussed by Haldane \cite{haldane83a,haldane83b}
 some time ago. Affleck and Haldane \cite{affleck85,affleck86,affleck87} related the O(3) non-linear $\sigma$-model with a topological
 term to the Wess-Zumino-Witten non-linear $\sigma$-model with topological coupling constant $k=1$, a conformal field theory that 
constitutes an attractive fixed point for antiferromagnetic Heisenberg chains with half-integer spin. Based on the conformal field-theory,
the critical behavior for $\Delta \neq 0$ was discussed by Affleck {\it et al.}
\cite{affleck89}, showing that the mass gap opens as 
$m \propto \Delta^{2/3}/|\ln \Delta|^{1/2}$. Numerical simulations with a cluster algorithm \cite{wiese95} confirmed 
later the correctness of the identification of the action (\ref{actioninnl}) with the WZW conformal field theory, and
 showed that for $\Delta \neq 0$, a gap opens. Hence, our path integral formulation allowed us to start from a dimerized 
phase and take into account the relevant long-wavelength fluctuations that describe the critical behavior of the system in 
the limit $\Delta \rightarrow 0$.

\subsection{Continuum limit with bond-fields}
As a first step, we have to identify as in the previous discussion, slow and fast components of the bond-fields, in order to allow for a gradient expansion. For that purpose, we use a staggered CP$^1$ representation, that was previously introduced for the t-J model \cite{falb08}. We discuss it shortly here, in order to allow for a self-contained presentation. 

\subsubsection{Staggered CP$^1$ representation \label{CP1}}
We consider now the fact that the spin-fields ${\vec S}_{(1)}$ and ${\vec S}_{(2)}$ on a bond are staggered with respect to each other, and introduce a CP$^1$ representation, such that
\bea
\label{cp1p}
S^a_{(1)} & = &\bar z^{(1)} \, \sigma^a \, z^{(1)} \; ,
\nonumber \\
S^a_{(2)} & = & z^{(2)}_\alpha \, \sigma^y_{\alpha \beta} \, 
\sigma^a_{\beta \gamma} \, \sigma^y_{\gamma \delta} \, z^{(2) *}_\delta \; ,
\eea
where ${\bar z}^{(i)} \, z^{(i)} = 1$, $i=1,2$.
On the other hand,  making the following replacements
\bea
\label{wztozAB}
\begin{array}{rclcrcl}
z_1 & \rightarrow & z^{(1)}_1 & \qquad \qquad & z_2 & \rightarrow & z^{(1)}_2
\\
\omega_1 & \rightarrow & -i z^{(2) *}_2 & \qquad \qquad & \omega_2 & \rightarrow  & i
 z^{(2) *}_1\;,
\end{array}
\eea 
in the expressions (\ref {Trans1}), the bond-fields can be written as follows:
\bea
\label{BondFromCPOneStaggered}
s & = & 
i \frac{\sqrt{2}}{2} \, {\bar z}^{(2)} z^{(1)}
 \; ,
\nonumber \\
t_a & = & 
i \frac{\sqrt{2}}{2}  \, {\bar z}^{(2)} \sigma^a z^{(1)} 
 \; .
\eea
From the relations above it is easy to show that the completeness relation
\bea
\label{CompletenessBondFields}
s^* s + t^*_a t^{}_a & = &
z^{(1) *}_\alpha z^{(1)}_\alpha  z^{(2) *}_\beta z^{(2)}_\beta = 1 \; ,
\eea
and that the constraints
\bea
\label{FurtherConstraintBondField}
ss - t^{}_a t^{}_a & = & s^* s^* - t^*_a t^*_a = 0 \; 
\eea
also hold.

The Berry phase  is given by
\bea
\label{BerryPhaseBondFields}
s^* \dot{s} + t_a^* {\dot{t}}^{}_a & = &
{\bar z}^{(1)} \partial_\tau z^{(1)}  - {\bar z}^{(2)} \partial_\tau z^{(2)}
\; .
\eea
This is the form of the Berry phase in the staggered CP$^1$ representation \cite{falb08}.

\subsubsection{Slow and fast components of the bond-fields}
For each bond we can define new fields
\bea
\label{newz}
{\tilde z} & = & \frac{1}{2} \left[ z^{(1)} + z^{(2)} \right] \; ,
\nonumber \\ 
a \zeta & = & \frac{1}{2} \left[ z^{(2)} - z^{(1)} \right] \; .
\eea
Due to the constraints satisfied by the fields $z^{(i)}$, we have
\bea
\left( \bar {\tilde z} + a \bar \zeta \right) 
\left( \tilde z + a \zeta \right) & = & 1 \; ,
\nonumber \\
\left( \bar {\tilde z} - a \bar \zeta \right) 
\left( \tilde z - a \zeta \right) & = & 1 \; ,
\eea
leading to constraints for the fields $\tilde z$ and $\zeta$,
\bea
\label{constzz}
\bar {\tilde z} \, \zeta + \bar \zeta \, \tilde z & = & 0 \; .
\\ 
\label{modzz}
\bar {\tilde z} \, \tilde z + a^2 \bar \zeta \, \zeta & = & 1 \; .
\eea
We then introduce new fields 
\bea
\label{newnewz}
\tilde z = z \sqrt{1 - a^2 \bar \zeta \zeta} \; ,
\eea 
such that the constraint (\ref{modzz}) goes over into
\bea
\label{modzzp}
\bar z z = 1 \; ,
\eea
and the constraint (\ref{constzz}) translates into
\bea
\label{constzzp}
\bar z \, \zeta + \bar \zeta \, z = 0 \; .
\eea
The fields $z$ and $\zeta$ defined above correspond to smooth configurations.
 
Next, we can express the bond-fields in terms of the fields $z$ and $\zeta$ as follows:
\bea
\label{BondFieldsFromZZeta}
s & = &  i \frac{\sqrt{2}}{2}
\left[1 + a \left(\bar \zeta z - \bar z \zeta \right) - 2 a^2 \bar \zeta \zeta \right] \; ,
\nonumber \\
t_a & = &  i \frac{\sqrt{2}}{2}
 \left[ \bar z \sigma^a z \left(1-a^2 \bar \zeta \zeta \right)
 + a \left(\bar \zeta \sigma^a z - \bar z \sigma^a \zeta \right)
 - a^2 \bar \zeta \sigma^a \zeta \right]
\; ,
\eea
where we keep contributions up to ${\cal O} \left(a^2\right)$.
Guided by the expression of $t_a$, we define the following vectors
\bea
\label{DefinitionsOfVectorFields}
\vec \Omega & = & \bar z \vec \sigma z \; ,
\nonumber \\
\vec L & = & \bar z \vec \sigma \zeta \; ,
\nonumber \\
{\vec L}^\dagger & = & \bar \zeta \vec \sigma z \; ,
\nonumber \\
\vec m & = & \bar \zeta \vec \sigma \zeta \; .
\eea
For later convenience, we define the following two vectors in addition to $\vec \Omega$ and $\vec m$
\bea
\label{DefinitionsOfLR}
{\vec L}_R & \equiv & \frac{1}{2} \left(\vec L + {\vec L}^\dagger \right)
= \frac{1}{2} \left(\bar z \vec \sigma \zeta + \bar \zeta \vec \sigma z \right)
\; ,
\\
\label{DefinitionsOfLI}
{\vec L}_I & \equiv & -\frac{i}{2} \left(\vec L - {\vec L}^\dagger \right)
= -\frac{i}{2} \left(\bar z \vec \sigma \zeta - \bar \zeta \vec \sigma z \right)
\; .
\eea
Further relations can be obtained by using that
\bea
\vec \Omega \cdot \vec L = \bar z \zeta \; , \quad
\vec \Omega \cdot {\vec L}^\dagger = \bar \zeta z \; ,
\eea
such that
\bea
\label{FromZetaZToOmegaDotLI}
\bar \zeta z - \bar z \zeta = \vec \Omega \cdot \left({\vec L}^\dagger - \vec L \right)
= - 2 i \vec \Omega \cdot {\vec L}_I
\; .
\eea
Using the constraint (\ref{constzzp}) it can be also shown that 
\bea
{\vec L}_I^2 & = & \bar \zeta \zeta \; .
\eea

With the relations above, we can rewrite (\ref{BondFieldsFromZZeta}) as follows
\bea
\label{BondFieldsWithEtaOmegaEllRandEllI}
s & = & \frac{i}{\sqrt{2}} \left[ 1 - 2 i a \vec \Omega \cdot {\vec L}_I
- 2 a^2 \left({\vec L}_I \right)^2  \right]
\; ,
\nonumber \\
t_a & = & \frac{i}{\sqrt{2}} \left[ \Omega_a - 2 i a L_{I a}
- a^2 \left( {\vec L}_I^2 \Omega_a + m_a \right)
\right]
\; .
\eea

Due to the constraints (\ref{modzzp}) and (\ref{constzzp}), the vector-fields fulfill the following conditions:
\bea
\label{ConstraintsVectorFields}
{\vec \Omega}^2 & = & 1 \; ,
\nonumber \\
\vec \Omega \cdot {\vec L}_R & = & 0 \; ,
\nonumber \\
{\vec L}_I \cdot {\vec L}_R & = & 0 \; ,
\nonumber \\
\vec m \cdot {\vec L}_R & = & 0 \; .
\eea
There we can see that the vector-fields $\vec \Omega$, ${\vec L}_I$, and $\vec m$ are coplanar, so that they can be described as linear combinations of two orthogonal fields in that plane. We choose them to be
$\vec \Omega$ and  $\vec \Omega \times {\vec L}_R$. Using the form of $\vec \Omega \times {\vec L}_R$ expressed in terms of the fields $z$ and $\zeta$ (eq.\ (\ref{DefinitionsOfVectorFields}) and eq.\ (\ref{DefinitionsOfLR})) we can write
\bea
\label{OmegaCrossLRp}
{\vec L}_I & = & 
\left(\vec \Omega \cdot {\vec L}_I \right) \vec \Omega -\left(\vec \Omega \times {\vec L}_R\right)  \; ,
\eea
and
\bea
\vec m & = & \left[2 \left( \vec \Omega \cdot {\vec L}_I \right)^2 - {\vec L}_I^2 \right] \vec \Omega
- 2 \left(\vec \Omega \cdot {\vec L}_I \right) \left(\vec \Omega \times {\vec L}_R \right) \; .
\eea
Inserting the expressions for ${\vec L}_I$ and $\vec m$ into (\ref{BondFieldsWithEtaOmegaEllRandEllI}), we have
\bea
\label{BondFieldsWithVectorFields}
s & = & \frac{i}{\sqrt{2}} \left[ 1 - 2 i a \, \varphi
- 2 a^2 \left(\varphi^2+{\vec L}_R^2 \right)  \right]
\; ,
\nonumber \\
\vec t & = & \frac{i}{\sqrt{2}} \left\{ \vec \Omega
- 2 i a \left[ \varphi \, \vec \Omega
- \left( \vec \Omega \times {\vec L}_R \right)
\right]
- 2 a^2 \varphi
\left[ \varphi \, \vec \Omega - \left(\vec \Omega \times {\vec L}_R \right) \right]
\right\}
\; ,
\eea
where we introduced $\varphi \equiv \vec \Omega \cdot {\vec L}_I$.
Here we see that using the fast and slow fields of the staggered CP$^1$ representation, we obtain an expansion of the bond-fields in terms of two vector- and one scalar-field. The constraints (\ref{CompletenessBondFields}) and (\ref {FurtherConstraintBondField}) on the bond-fields are fulfilled by imposing the conditions (\ref{ConstraintsVectorFields}), such that there is no condition on the scalar field 
$\varphi$ at this stage.

Recalling that the bond-fields are related to the SO(4) fields as follows:
\bea
\label{FromSO4ToBonds}
{\cal S}_a & = & s^* t^{}_a + t^*_a s \; ,
\nonumber \\
{\cal T}_a & = & - i \varepsilon_{abc} t^*_b t^{}_c \; ,
\eea
we can relate the SO(4) vector-fields with the fields entering (\ref{BondFieldsWithVectorFields}), arriving at
\bea
\label{SO4withVectorFieldsFromBondFields}
\vec {\cal S} & = &
\vec \Omega - 2 a^2 \bigg[{\vec L}_R^2 \,
\vec \Omega
+ \varphi  \left(\vec \Omega \times {\vec L}_R \right)
\bigg]
\; ,
\nonumber \\
\vec {\cal T} & = & 
- 2 a {\vec L}_R \; .
\eea
One can easily see that the fields above fulfill the constraints obeyed by the fields
$\vec {\cal S}$ and $\vec {\cal T}$ up to ${\cal O} \left(a^2 \right)$.

\subsubsection{Constraints and measure}\label{Constraints and measure}
The expansion (\ref{BondFieldsWithVectorFields}) of the bond-fields in fast and slow components involves only seven components instead of eight, as originally introduced. Taking into account the first two
constraints in (\ref{ConstraintsVectorFields}), we are left with five fields instead of the four required to describe the degrees of freedom on a dimer formed by two $S=1/2$ spins. Therefore, before passing to a path integral in $\varphi$, $\vec \Omega$, and ${\vec L}_R$, a gauge fixing condition is still necessary. Here we choose the gauge fixing $t^*_3 + t^{}_3 =0$. Once we proposed an expansion of the bond-fields in terms of fast and slow variables, we consider first the constraints on those variables together with the change of measure due to the transformation to these new variables.

The measure and constraints corresponding to the general gauge-fixing  were obtained in Sec.\ \ref{FJBondFields} (eq. \ref{PIst0}). For the particular gauge $\varphi_{4}=t^*_3 + t_3 $ the measure can be written as follows:
\bea
{\cal DM} & = &
{\cal D}s^* \, {\cal D}s \,
{\cal D}t^*_{a} \, {\cal D}t^{}_a \,
(t^*_3 - t^{}_3) \,
\delta \left(t^*_3 + t^{}_3  \right) \,
\nonumber \\ & & \times
\delta \left(s^* s + t^*_a t^{}_a-1 \right) \,
\delta \left(s^* s^*-t^*_a t^*_a\right) \,
\delta \left(s s - t_a t_a \right) \; .
\eea
Since we have in total seven variables given by $\varphi$, $\vec \Omega$, and ${\vec L}_R$, we can eliminate one of the variables by integrating over $t^*_3$ and imposing the constraint due to gauge fixing. Then, we have
\bea
\label{MeasureForChange}
{\cal DM} & \rightarrow &
{\cal D}s \,
{\cal D}t^*_{a} \, {\cal D}t^{}_a \,
t_3 \,
\delta \left(s^* s + t^*_1 t^{}_1+ t^*_2 t^{}_2 - t^2_3-1 \right)
\nonumber \\ & & 
\times
\delta \left(s^* s^* -t^*_1 t^{}_1 - t^*_2 t^{}_2 - t^2_3\right)\,
\delta \left(s s - t_a t_a \right) \; .
\eea
The arguments of the $\delta$-functions with the new variables look as follows
\bea
\lefteqn{
\delta \left(s^* s + t^*_1 t^{}_1+ t^*_2 t^{}_2 - t^2_3-1 \right)
}
\nonumber \\
& = &
\delta\left\{
\frac{{\vec \Omega}^2-1}{2}
- 2 i a \Omega_3 \left[ \varphi \Omega_3
- \left(\Omega_1 L_{R 2} - \Omega_2 L_{R 1} \right) \right]\right.
\nonumber \\ & &\left. \quad
- 2 a^2 \bigg [
\left( \vec \Omega \cdot {\vec L}_R \right)^2
+ {\vec L}_R^2 \left(1- {\vec \Omega}^2 \right)
+ 2 \left[ \varphi \Omega_3
- \left(\Omega_1 L_{R 2} - \Omega_2 L_{R 1} \right)
\right]^2
\bigg]\right\} \; ,
\nonumber\\ 
\lefteqn{
\delta \left(s^* s^* -t^*_1 t^{}_1 - t^*_2 t^{}_2 - t^2_3\right)
}
\nonumber \\
 & = & \delta \bigg\{
\frac{{\vec \Omega}^2-1}{2}
- 2 i a \left\{
\left(1-{\vec \Omega}^2 \right) \varphi
+2 \Omega_3 \left[
\varphi \Omega_3
- \left(\Omega_1 L_{R 2} - \Omega_2 L_{R 1} \right)
\right]
\right\}
\nonumber \\ & & \quad 
+ 2 a^2 \left[
\left(\vec \Omega \cdot {\vec L}_R \right)^2
- \left({\vec \Omega }^2 -1 \right)
\left(
{\vec L}^2_R + 2 \varphi \right)
\right]\bigg \} \; ,
\nonumber \\
\lefteqn{
\delta \left(s s - t_a t_a \right)
}
\nonumber \\
& = &\delta \bigg \{
\frac{{\vec \Omega}^2-1}{2}
- 2 i a \varphi \left({\vec \Omega}^2 -1 \right)
+ 2 a^2 \left[\left(\vec \Omega \cdot {\vec L}_R \right)^2
- \left({\vec L}_R^2 + 2 \varphi^2 \right)
\left({\vec \Omega}^2 -1 \right)
\right]\bigg \} \; .
\eea
We can transform these constraints onto constraints on the new fields:
\bea
\lefteqn{\delta \left(s^* s + t^*_1 t^{}_1+ t^*_2 t^{}_2 - t^2_3-1 \right)\;\delta \left(s^* s^* -t^*_1 t^{}_1 - t^*_2 t^{}_2 - t^2_3\right) \;
\delta \left(s s - t_a t_a \right)
}
\nonumber\\ && \qquad \qquad
 =  \frac{1}{J_\delta} \, \delta \left( {\vec \Omega}^2 -1 \right)
\;
\delta \Big[ \varphi \Omega_3
- \left(\Omega_1 L_{R 2} - \Omega_2 L_{R 1} \right) \Big] \,
\delta \left[\left(\vec \Omega \cdot {\vec L}_R \right)^2 \right] \; ,
\eea
where the inverse of the Jacobian for the transformation is up to
${\cal O} \left(a^2 \right)$ 
\bea
J^{-1}_\delta & = & \frac{1}{8 \Omega_3} + i a \frac{\varphi}{2 \Omega_3}
- a^2
\frac{\varphi \left(2 \varphi \Omega_3 + \Omega_1 L_{R 2} - \Omega_2 L_{R 1}\right)}
{2 \Omega_3^2}
\; .
\eea
On the other hand, the Jacobian for the transformation to the new fields is up to
${\cal O} \left(a^2 \right)$ 
\bea
J & = & 16 \sqrt{2} \, \left(\vec \Omega \cdot {\vec L}_R \right)
\Big\{ \Omega_3
- 2 i a \left[\varphi \Omega_3 + \left(\Omega_1 L_{R 2} - \Omega_2 L_{R 1} \right)\right]
\nonumber \\ & & \qquad \qquad \qquad \quad
- 2 a^2 \varphi
\left[\varphi \Omega_3 + \left(\Omega_1 L_{R 2} - \Omega_2 L_{R 1} \right)\right]
\Big\} \; .
\eea
The first factor is cancelled in going from
$\delta \left[\left(\vec \Omega \cdot {\vec L}_R \right)^2 \right]$ to
$\delta \left(\vec \Omega \cdot {\vec L}_R \right)$. Finally, the measure
in (\ref{MeasureForChange}) is now
\bea
\label{MeasureWithPhiAndVectors}
{\cal DM} & \propto  &
 {\cal D} \varphi \,
{\cal D} \vec \Omega \, {\cal D} {\vec L}_R \,
\delta \left( {\vec \Omega}^2 -1 \right)
\delta \left[ \varphi
- \frac{\left(\Omega_1 L_{R 2} - \Omega_2 L_{R 1} \right)}{\Omega_3} \right] \,
\delta \left(\vec \Omega \cdot {\vec L}_R \right).
\eea

\subsubsection{Field-theory for the spin-Peierls model}
Changing notation ${\vec L}_R \rightarrow \vec L$, for brevity, we just express the SO(4)-fields in terms of
 $\varphi$, $\vec \Omega$, and $\vec L$ using (\ref{FromSO4ToBonds}), where keeping terms up to 
${\cal O} \left(a^2 \right)$ leads to (\ref{SO4withVectorFieldsFromBondFields}). Going back to (\ref{HamiltonianSPWithSO4}), 
we can express the Hamiltonian with the new fields, leaving aside constant terms,
\bea
\label{HamiltonianSPWithExpandedBond-Fields}
H_{SP} & = &
\int {\rm d} x \,\frac{J\,a}{4}\; \Bigg[ 16 {\vec L}^2+(1-\Delta)\left(
\frac{1}{2} \;
\partial_x  {\vec \Omega} \cdot \partial_x  {\vec \Omega}
- 4\,\partial_x {\vec \Omega} \cdot {\vec L}
\right)  \Bigg]
\; .
\eea
Next we consider the Berry-phase as given by (\ref{BerryPhaseBondFields}), where we need to consider only 
contribution up to ${\cal O} (a) $, and discard total time derivatives,
\bea
S_B & = & \int {\rm d} \tau \sum_i \left(s^* \dot{s} + t_a^* {\dot{t}}_a \right)
=
-2 i \int {\rm d} \tau \, {\rm d} x \,
{\vec L} \cdot \left({\vec \Omega} \times \partial_\tau {\vec \Omega} \right)
\; .
\eea
Finally, the total action is given by
\bea
\label{ActionSPFromBondFields}
S & = &  \int {\rm d} \tau \, {\rm d} x
\Bigg\{
-2 i {\vec L} \cdot \left({\vec \Omega} \times \partial_\tau {\vec \Omega}  \right)
\nonumber \\ & & \qquad \qquad
+ \frac{J\,a}{4}\; \Bigg[ 16 {\vec L}^2+(1-\Delta)\left(
\frac{1}{2} \;
\partial_x  {\vec \Omega} \cdot \partial_x  {\vec \Omega}
- 4\,\partial_x {\vec \Omega} \cdot {\vec L}
\right)  \Bigg]
\Bigg\} \; ,
\eea
with the partition function
\bea
\label{PartitionFunctionSPFromBondFields}
Z & = & \int {\cal D} \varphi \,
{\cal D} \vec \Omega \, {\cal D} {\vec L} \,
\delta \left( {\vec \Omega}^2 -1 \right) \,
\delta \left(\vec \Omega \cdot {\vec L} \right)
\;\delta \left[ \varphi
- \frac{\left(\Omega_1 L_2 - \Omega_2 L_1 \right)}{\Omega_3} \right] \,
{\rm e}^{-S} \; .
\eea
Therefore, we can trivially integrate out $\varphi$ and obtain
\bea
Z & = & \int {\cal D} \vec \Omega \, {\cal D} {\vec L} \,
\delta \left( {\vec \Omega}^2 -1 \right) \,
\delta \left(\vec \Omega \cdot {\vec L} \right) \,
{\rm e}^{-S} \; ,
\eea
with the action given by (\ref{ActionSPFromBondFields}). As previously done, we can integrate out $\vec L$, or what is equivalent, we solve for the saddle point for $\vec L$ as done in subsec.\ \ref{SO(4)formulation1}, arriving at
\bea
L_a & = & \frac{(1-\Delta)}{8} \partial_x \Omega_a
- \frac{i}{4 J_1 a} \varepsilon^{abc} \Omega_b \partial_\tau \Omega_c
\; ,
\eea
leading to the effective action
\bea
\label{EffectiveActionGaplessPhase}
\tilde S & = &  \int {\rm d} \tau \, {\rm d} x \,
\left[ \frac{1}{4J a} \left( \partial_\tau \vec \Omega \right)^2
+ \frac{J\, a}{16} (1-\Delta^{2}) \left(\partial_x  {\vec \Omega} \right)^2
\right]
\nonumber \\ & &
+ \frac{i}{4} \,(1-\Delta)\,
\int {\rm d} \tau \, {\rm d} x \, \vec \Omega
\cdot \left(\partial_\tau \vec \Omega \times \partial_x \vec \Omega \right) \; ,
\eea
that, as expected, coincides with the action (\ref{actioninnl}).

\section{Summary}
We have presented path integral formulations for dimerized quantum antiferromagnets both for the SO(4)-algebra 
obeyed by the operators on a bond, as well as for bond-operators \cite{sachdev90}. We used the formalism introduced 
by Faddeev and Jackiw for the quantization of constrained systems \cite{Faddeev88,Govaerts90,Barcelo92, montani93}, 
where the presence of constraints is revealed by singular modes of the symplectic matrix resulting from the canonical 
form in the Lagrangian. As opposed to the Dirac treatment of constrained systems \cite{Dirac64}, no distinction among 
primary and secondary, first class or second class constraints is needed, but an iterative incorporation of constraints 
determined by the singular modes, into the canonical form. For the SO(4)-formulation, the FJ-formalism ensures that the
 basic brackets of the fields corresponds to those of the generators of the algebra. On the other hand, bond-operators 
posses a gauge degree of freedom, such that depending on the form of the gauge-fixing, different forms for the basic 
brackets may result, a fact that is also known from the quantization of slave-particle formulations of the t-J model \cite{leguillou95}.           
As an application for both formulations, we considered the spin-Peierls model in one dimension, where
we obtained the corresponding field-theories. This model, albeit simple and well known, was chosen since the gapless
 phase is realized due to the presence of a topological term. The present treatment shows how, by taking into account
 properly fluctuations in a gradient expansion, this non-trivial information is encoded in the fields appropriate for 
dimerized states.  Since the path integral formulation presented here is not restricted by dimensionality,
we expect this treatment to be useful in dealing with dimerized quantum antiferromagnets in higher dimension, 
as those mentioned in the Introduction.

\vspace*{0.5cm}
\par
\noindent
{\bf Acknowledgements}
\par
\noindent
We are grateful to DAAD through the Program PROALAR for financial support of our collaboration. We also acknowledge partial support by the DFG through SFB/TRR 21. A. M. is grateful to the Aspen Center of Physics for hospitality.

\appendix
\section {Mapping between the path integral formulations for SO(4) and bond fields }
In the present appendix, we obtain the path integral representation for the SO(4) generators  (eqs.\ (\ref{Z_STau}) and (\ref{ActionTauS})) from the path integral representation for the bond fields .
The starting point is eq.\ (\ref{PIst0})
\begin{eqnarray}
\label{PIst0p}
{\mathcal{Z}} = \int\,{\mathcal{D}}s^{*}\,{\mathcal{D}}s
\,{\mathcal{D}}t^{*}_{a}\,{\mathcal{D}}t_{a} \,(det\;\left[M_{AB}\right])^{1/2} \,\delta[\varphi_{4}]\,\delta[\varphi_{1}] \, \delta[\varphi_{2}] \, \delta[\varphi_{3}] \, 
e^{-S} \; ,
\end{eqnarray}
where the action is given by eq.\ (\ref{ActionBondFiledsGeneral}),
with the constraints (\ref{newvinvarphi1}), (\ref{newvin2}),
and the gauge fixing condition
$\varphi_{4}=t^{*}_{3}+t_{3}$, as in Sec.\ (\ref{Constraints and measure}).
Then, from the equation (\ref{detst}) we have
\begin{eqnarray}
\left( det\;\left[ M_{AB}\right]\right) ^{1/2}& = &
4\;\left( t_{3}-t^{*}_{3}\right) .
\end{eqnarray}
Next, we introduced in (\ref{PIst0p}) the SO(4) fields given by  (\ref{FromSO4ToBonds}) through the identity  
\begin{eqnarray}\label{unoSTau}
 {\mathbf{1}} & = & \int\;{\mathcal{D}} \mathcal{S}_{1}\;{\mathcal{D}} \mathcal{S}_{2}\;{\mathcal{D}} \mathcal{S}_{3}\;{\mathcal{D}} \mathcal{T}_{1}
\;{\mathcal{D}} \mathcal{T}_{2}
\;{\mathcal{D}} \mathcal{T}_{3}
\; {\cal P} 
\; \delta[ \mathcal{T}_{3}+i(t^{*}_{1}\;t_{2}-t^{*}_{2}\;t_{1})] 
\; ,
 \end{eqnarray} 
 where ${\cal P}$ is a product of the remaining $\delta$-functions that relate the SO(4) with bond-fields:  
 \begin{eqnarray}
{\cal P} & = & \delta[\mathcal{S}_{1}-(s^{*}t_{1}+t^{*}_{1}\;s) ]
\;\delta[\mathcal{S}_{2}-(s^{*}t_{2}+t^{*}_{2}\;s)] \;
\delta[\mathcal{S}_{3}-(s^{*}t_{3}+t^{*}_{3}\;s) ]
\nonumber\\ & & \times
\;\delta[\mathcal{T}_{1}+i(t^{*}_{2}\;t_{3}-t^{*}_{3}\;t_{2})] \; 
\delta[\mathcal{T}_{2}+i(t^{*}_{3}\;t_{1}-t^{*}_{1}\;t_{3})] 
\; .
 \end{eqnarray}
It is possible to show that
\begin{eqnarray}
\label{STau_st}
\lefteqn{
{\cal P} \; 
\;\delta[s\,s-t_{\alpha}\;t_{\alpha}]
\;\delta[s^{*}\,s^{*}-t^{*}_{\alpha}\;t^{*}_{\alpha}]
}
\nonumber \\ &  & =
\frac{1}{J}\;\delta\left[s^{*}-\left(\frac{\mathcal{S}_{3}}{2\,t_{3}}+i\,t_{3}\,\frac{\mathcal{T}_{2}\,\mathcal{S}_{1}-\mathcal{T}_{1}\,\mathcal{S}_{2}}{\mathcal{S}^{2}_{3}+\,\mathcal{T}^{2}_{2}+\mathcal{T}^{2}_{1}}\right)\right] 
\delta\left[s+\left(\frac{\mathcal{S}_{3}}{2\,t_{3}}-
i\,t_{3}\,\frac{\mathcal{T}_{2}\,\mathcal{S}_{1}-\mathcal{T}_{1}\,
\mathcal{S}_{2}}{\mathcal{S}^{2}_{3}
+\,\mathcal{T}^{2}_{2}+\mathcal{T}^{2}_{1}}\right) \right]\;
\nonumber\\ & & \quad \times
\delta\left[t^{*}_{1}+\left( \frac{t_{3}\,\mathcal{S}_{1}-i\,s^{*}\,\mathcal{T}_{2}}{\mathcal{S}_{3}}\right)\right]\;
\delta\left[t_{1}-\left( \frac{t_{3}\,\mathcal{S}_{1}-i\,s\,\mathcal{T}_{2}}{\mathcal{S}_{3}}\right)\right ]\;
\delta\left[t^{*}_{2}+\left( \frac{t_{3}\,\mathcal{S}_{2}-i\,s^{*}\,\mathcal{T}_{1}}{\mathcal{S}_{3}}\right)\right]\;\nonumber\\ && \quad \times
\delta\left[t_{2}-\left( \frac{t_{3}\,\mathcal{S}_{2}-i\,s\,\mathcal{T}_{1}}{\mathcal{S}_{3}}\right) \right] \;
\delta\left[t_{3}-i\,\sqrt{\frac{\mathcal{S}_{3}\,(\mathcal{S}^{2}_{3}
+\,\mathcal{T}^{2}_{2}+\mathcal{T}^{2}_{1})}
{\sqrt{(\mathcal{S}_{1}\,\mathcal{T}_{1}+\mathcal{S}_{2}\,\mathcal{T}_{2})^{2}
+\mathcal{S}^{2}_{3}\,
(\vec{\mathcal{S}}^{2}+\mathcal{T}^{2}_{1}+\mathcal{T}^{2}_{2})}}}\right ]
\; ,
\end{eqnarray}
where ${J}\, = -4 \mathcal{S}_{3} \left( t_{3}-t^{*}_{3}\right)$ is the  Jacobian of the change of variables. 
Using (\ref{STau_st}), and after integrating over the bond-fields, we arrive at
\begin{eqnarray}
\label{Z_STauAp}
 \mathcal{Z }& = & \int {\cal D} \mathbf{\mathcal T}
\, {\cal D} \mathbf{\mathcal S} \, \delta \left( \mathbf{\mathcal T} \cdot
\mathbf{\mathcal S} \right) \, \delta \left({\mathbf{\mathcal T}}^2 + {\mathbf{\mathcal S}}^2 - 1\right) \,
 \; e^{-S
%  i\int\;L(\mathbf{\mathcal T}, \,\mathbf{\mathcal S} )\;dt
} \; ,
 \end{eqnarray}
 where the Berry phase is given by
 \begin{eqnarray}
 \label{BP1}
(s^* \dot{s} + t^*_a \dot{t}^{}_{a})
%\nonumber\\
& \rightarrow &
\left[ -\frac{\mathcal{S}_{1}}{\mathcal{S}_{3}}\dot{\mathcal{T}_{2}}+\frac{\mathcal{S}_{2}}{\mathcal{S}_{3}}\dot{\mathcal{T}_{1}}+\frac{1}{2}\;\frac{(\mathcal{T}_{2}\mathcal{S}_{1}-\mathcal{T}_{1}\mathcal{S}_{2})}{\mathcal{S}_{3}(\mathcal{S}^{2}_{3}+\,\mathcal{T}^{2}_{2}+\mathcal{T}^{2}_{1})}\;\frac{d}{d\tau}\left(\mathcal{S}^{2}_{3}+\,\mathcal{T}^{2}_{2}+\mathcal{T}^{2}_{1}\right) \right]
\; ,
%&&
%+ derivadas\;\;totales
\end{eqnarray}
resulting in the following contributions after simplifying the expressions using the constraints of the SO(4) fields,
\begin{eqnarray}
\label{coefBP}
\mathcal{A}^\mathcal{T}_{1} & = &
\frac{\mathcal{S}_{2}\mathcal{S}_{3}-\mathcal{T}_{2}\mathcal{T}_{3}}{(\mathcal{S}^{2}_{3}+\,\mathcal{T}^{2}_{2}+\mathcal{T}^{2}_{1})} \; ,
\nonumber\\
\mathcal{A}^\mathcal{T}_{2} & = & \frac{-\mathcal{S}_{1}\mathcal{S}_{3}+\mathcal{T}_{1}\mathcal{T}_{3}}{(\mathcal{S}^{2}_{3}+\,\mathcal{T}^{2}_{2}+\mathcal{T}^{2}_{1})} \; ,
\nonumber\\
\mathcal{A}^\mathcal{T}_{3} & = &
%0\nonumber\\
\mathcal{A}^\mathcal{S}_{1} =
%&0\nonumber\\
\mathcal{A}^\mathcal{S}_{2} = 0 \; ,
\nonumber\\
\mathcal{A}^\mathcal{S}_{3}&=&\frac{(\mathcal{T}_{2}\mathcal{S}_{1}-\mathcal{T}_{1}\mathcal{S}_{2})}{(\mathcal{S}^{2}_{3}+\,\mathcal{T}^{2}_{2}+\mathcal{T}^{2}_{1})} \; ,
 \end{eqnarray}
where the curl of the fields above only partially fulfill eqs.\ (\ref{ecu_finales2}) and (\ref{Solution1}). However, they comply with eqs.\  (\ref{ecu_finales1}). Using the expressions (\ref{coefBP}), and the expansions of the fields $\vec {\cal S}$ and $\vec {\cal T}$ from eq.\ (\ref{FieldExpansion}), we arrive at the following expression for the Berry phase:
\bea
{\vec {\cal A}}^{\cal T} \cdot \partial_\tau
\vec {\cal T} + {\vec {\cal A}}^{\cal S} \cdot \partial_\tau \vec {\cal S} 
& = &
-i \frac{1}{n_3} \left(\ell_1 \partial_\tau n_2 + \ell_1 \partial_\tau n_2 \right) + {\cal O} \left(a^3 \right)
\; .
\eea
It is straightforward to show using the fact that ${\vec n}^2 = 1$ and $\vec n \cdot \vec \ell = 0$, 
that the expression above is equal to the integrand in eq.\ (\ref{BerryPhasep}).

\end{document}